\documentclass[prd,onecolumn,nofootinbib,aps,tightenlines,
preprintnumbers]{revtex4-1}

\def\be{\begin{equation}}
\def\ee{\end{equation}}
\def\bea{\begin{eqnarray}}
\def\eea{\end{eqnarray}}

\newcommand{\del}{\partial}

\usepackage{framed}
\usepackage[compat=1.0.0]{tikz-feynman}
\usepackage{multirow}
\usepackage{braket}
\usepackage{color}
\usepackage{amsfonts}
\usepackage{amssymb}
\usepackage{amsthm}
\usepackage{slashed}
\usepackage{bm}
\usepackage[makeroom]{cancel}
\usepackage{amsmath, mathtools}
\usepackage{framed}

\usepackage{graphicx}

\usepackage[margin=0.7in]{geometry}

\newcommand{\Lagr}{\mathcal{L}}

\newcommand{\units}[1]{\ensuremath{\mathrm{#1}}}

\newcommand{\mev}{\units{MeV}}
\newcommand{\gev}{\units{GeV}}

\newcommand{\cm}{\units{cm}}

\newcommand{\s}{\units{s}}
\newcommand{\sr}{\units{sr}}
\newcommand{\pb}{\text{pb}}
\newcommand{\yr}{\text{yr}}

\begin{document}

\preprint{\hbox{PREPRINT UH511-1302-2019, UCI-TR-2019-04}  }

\title{
Dark Matter Through the Quark Vector Current Portal
}
\author{Dillon Berger}
\affiliation{Department of Physics and Astronomy, University of
California, Irvine,
  CA 92697, USA}
\author{Jason Kumar}
\affiliation{Department of Physics and Astronomy, University
of Hawai'i,
  Honolulu, HI 96822, USA}
\author{Arvind Rajaraman}
\affiliation{Department of Physics and Astronomy, University of
California, Irvine,
  CA 92697, USA}

\begin{abstract}
We consider models of light dark matter coupled to quarks through a vector current interaction.
For low energies, these models must be treated through the effective couplings to mesons, which
are implemented here through the chiral Lagrangian. We find the rates of dark matter annihilation
and decay to
the light mesons, and find the expected photon spectrum  from the decay of the hadrons.
We compare to current and future observations, and show
that there is a significant discovery reach for these models.
\end{abstract}
\maketitle

\section{Introduction}

Recently, there has been significant interest in models of dark matter (DM) in
which the dark matter particle
has a mass $m_X \lesssim {\cal O}(\gev)$.
These models
can evade the tight constraints on dark matter placed by direct detection
experiments, since these experiments typically lose
sensitivity at low mass.  If dark matter with
$m_X \lesssim {\cal O}(\gev)$ annihilates or decays
in the cosmos,
the photons produced will tend to lie in the current ``MeV-gap" in
observational sensitivity, but a variety of
new instruments (such as e-ASTROGAM~\cite{DeAngelis:2017gra},
AMEGO~\cite{Caputo:2017sjw} and APT~\cite{APT}) are being developed to fill this gap.  Such
instruments would be
well-positioned for indirect detection searches for MeV-range dark matter.

There has been particular recent interest in MeV-range dark matter which
couples largely to light
quarks~\cite{Boddy:2015efa,Boddy:2015fsa,Boddy:2016fds,Bartels:2017dpb,Cata:2017jar,
Dutra:2018gmv,Kumar:2018heq}.  The reason is
because the hadronic final states which can be produced at such small
center-of-mass energies are largely constrained
by kinematics and symmetry.  Moreover, several of the accessible hadrons,
such as $\pi^0$ and $\eta$, produce striking
photon signals when they decay.  This scenario is thus particularly
appealing from the point of view of indirect detection.

Recent work has considered the case where dark matter couples to either
scalar,  pseudoscalar, or axial-vector quark
bilinears~\cite{Kumar:2018heq}.
But if the dark matter couples to a quark vector current, then the leading accessible
final state (at low center-of-mass energy)
is  $\pi^+ \pi^-$, whose decays produce few photons, making this case difficult to
probe.  In this work, we revisit this case at slightly
higher center-of-mass energy ($\sqrt{s} \gtrsim 1~\gev$), where new final states are
allowed.  We determine the photon spectrum which
will be produced for a variety of choices of the flavor structure of DM-Standard
Model (SM) interactions, and determine the sensitivity of proposed
experiments.

We assume
that electroweak couplings are only relevant for the decays of hadrons produced by
dark matter annihilation/decay.  As described in~\cite{Kumar:2018heq} (see also~\cite{Choudhury:2019tss}),
DM-SM interactions can then be understood using
chiral perturbation theory, where dark matter
is introduced as a spurion which breaks Standard Model flavor symmetries.  We will
be interested in the case where dark matter
appears as a vector spurion.  The chiral Lagrangian used for the analysis
in~\cite{Kumar:2018heq} only involved the pseudoscalar meson
octet.  But since this work will consider a higher mass range, we will find
that we must also introduce the vector meson octet,
following the approach of~\cite{Terschlusen:2012xw}.  We will find that the
dominant photon signal arises from the production
of $\pi^0$, either directly or from the decays of $K_L$, $K_S$, $K^\pm$, $\rho$ or $\omega$.

The plan of this paper is as follows.  In Section II, we will describe the application of
chiral perturbation theory to
the interaction of low-mass dark matter with quark vector currents.  In Section III, we
will describe the generation of
the photon spectrum through primary and secondary decays of the hadronic final state
particles.  In Section IV, we will
present the sensitivities which can be expected from proposed experiments.  We conclude
with a discussion of our results.

\section{The Application of Chiral Perturbation Theory to Dark Matter Interactions With
Vector Currents}

We will consider two models
in which dark matter either decays or annihilates through a coupling to a vector quark current.
In the first model of
dark matter decay, a single spin-1 dark matter particle ($X_\mu$) couples to the vector quark
current and can decay to Standard Model particles. For this model,
\bea
{\cal L}_{int} &=&  g\sum_q \alpha_q X^\mu \bar{q}\gamma_\mu q ,
\label{eq:IntLagrangian}
\eea
In the second model,
we take  dark matter to  be a Dirac fermion ($\chi$)
which couples to quarks through a vector-vector interaction.
In this model we take
\bea
{\cal L}_{int} &=&  \sum_q {\alpha_q \over \Lambda^2}\bar{\chi}\gamma^\mu\chi
 \bar{q}\gamma_\mu q ,
\label{eq:IntLagrangian2}
\eea
For the application to chiral perturbation theory, it is useful to consider these interactions
as couplings of the quark vector current to a spurion $v^\mu_q$
where,
\bea
v^\mu_q &=& g\alpha_q X^\mu, ~{\rm or}~ {\alpha_q \over \Lambda^2}\bar{\chi}\gamma^\mu\chi.
\eea
Note that this interaction generically breaks the $SU(3)_L \times SU(3)_R$ flavor symmetries of
low-energy QCD.

We will take the mass of the dark matter to be in the GeV range; we shall be more specific
shortly. In this range the decay/annihilation products cannot be treated as weakly coupled propagating
quarks.
We assume that the dominant final states produced by low-mass dark matter interactions
with quarks are
hadronic, and that primary interactions which scale as $\alpha_{EM}$ or $s G_F$ are
negligible.  Since
the coupling of dark matter to light hadrons is governed by QCD and the dark
matter-quark current contact
interaction, we can directly express the coupling of dark matter to light mesons
using a chiral Lagrangian
in which dark matter appears as a spurion which breaks Standard Model flavor
symmetries.

This approach was followed in~\cite{Kumar:2018heq}, under the assumption that
$\sqrt{s} \lesssim 2m_{K^\pm}$.
In this case, the only accessible hadrons are $\pi^0$, $\pi^\pm$ and $\eta$, and one
can describe DM-SM
interactions using a chiral Lagrangian involving only the spurions and the pseudoscalar
meson octet.  But in
the case in which dark matter only interacted with quark vector currents, it was found
that the only accessible
two-body final state
was $\pi^+ \pi^-$.  Since
the decays of charged
pions produce very few photons, this channel is not useful for the purpose of indirect
detection with gamma-ray
telescopes.

We now consider the case in which the energy reach is extended to
$\sqrt{s} \lesssim  1.15~\gev$; if we assume
that the mass of the lightest glueball state is approximately $1~\gev$, then this
upper cutoff represents the
energy scale at which glueball states  can be produced along with neutral
pions.  A treatment of glueball emission
is beyond the scope of this work, and although the results we find may be qualitatively
correct even for $\sqrt{s}
\gtrsim 1.15~\gev$, we will not study this mass range further.

The amplitude for producing any of the allowed final states from a dark matter initial state can be
calculated using the chiral Lagrangian (these states are restricted by symmetry considerations; see
Appendix \ref{symmetries}).
We first write the chiral Lagrangian for the
pseudoscalar meson octet  to lowest order in the $p^2$ expansion.  As the flavor
symmetries of low-energy QCD are broken by the insertion of the $v_q^\mu$ in the fundamental
Lagrangian through eq.~\ref{eq:IntLagrangian}, the $v_q^\mu$ must also appear in the chiral Lagrangian
as spurions which break the flavor symmetries.  The form of these interactions is determined at
this order
by symmetry considerations, and chiral Lagrangian can be expressed (see~\cite{CHPTReviews}) as
\bea
{\cal L}_{\Phi}
&=& {F^2\over 4}Tr\left[(\del_\mu U - iv_\mu U + iU v_\mu\right)
\left( \del^\mu U^\dagger + iU^\dagger v_\mu-iv_\mu U^\dagger )\right] ,
\eea
where
\bea
U &\equiv& \exp [\imath \sqrt{2} \Phi / F] ,
\nonumber\\
\Phi &\equiv& \left(
                \begin{array}{ccc}
                  \frac{\pi^0}{\sqrt{2}} + \frac{\eta_8}{\sqrt{6}} & \pi^+ & K^+ \\
                  \pi^- & -\frac{\pi^0}{\sqrt{2}} + \frac{\eta_8}{\sqrt{6}} & K^0 \\
                  K^- & \bar K^0 & -\frac{2\eta_8}{\sqrt{6}} \\
                \end{array}
              \right) ,
\nonumber\\
v^\mu &\equiv&
\left(
                \begin{array}{ccc}
                  v^\mu_u & 0 & 0 \\
                  0 & v^\mu_d & 0 \\
                  0& 0 & v^\mu_s\\
                \end{array}
              \right) .
\label{eq:PhiLag}
\eea
The pion decay constant is $F \sim 92~\mev$.
To a good
approximation,
$\eta_8$ can be equated with the physical $\eta$ meson, and we do so henceforth.

It is useful to parameterize the flavor structure of $v_\mu$ in terms of its transformation properties under
the isospin subgroup of $SU(3)_L \times SU(3)_R$.  In particular, $v_\mu$ can be expanded in terms of two
isospin singlet components $v^{1Is}$ and $v^{s2I}$, and a component $v^{tI}$ with quantum numbers $I=1$, $I_3=0$:
\bea
v^{s1I} &\propto& \left(
                \begin{array}{ccc}
                  \frac{1}{\sqrt{2}} & 0 & 0 \\
                  0 & \frac{1}{\sqrt{2}} & 0 \\
                  0& 0 & 0 \\
                \end{array}
              \right) ,
\nonumber\\
v^{s2I} &\propto& \left(
                \begin{array}{ccc}
                  0 & 0 & 0 \\
                  0 & 0 & 0 \\
                  0& 0 & 1 \\
                \end{array}
              \right) ,
\nonumber\\
v^{tI}  &\propto& \left(
                 \begin{array}{ccc}
                  \frac{1}{\sqrt{2}} & 0 & 0 \\
                  0 & -\frac{1}{\sqrt{2}} & 0 \\
                  0& 0 & 0\\
                \end{array}
              \right) .
\eea
The isospin quantum numbers of the final state should be the same as those of the vector spurion.

The vector spurion
only couples to an even number of mesons in ${\cal L}_{\Phi}$; since we are only interested in two-body
final states, the only relevant terms in eq.~\ref{eq:PhiLag} are the following contact interactions
\bea
{\cal L}_{\rm contact}= \imath \bigg{\{}  (v^\mu_d - v^\mu_s)
\bar{K}^0\del_\mu K^0
+(v^\mu_s - v^\mu_u)
K^+ \del_\mu K^-
+ (v^\mu_d - v^\mu_u)
\pi^+\del_\mu\pi^- - h.c \bigg{\}}.
\label{eq:PSMesonCoupling}
\eea
Note that, these interactions respect isospin, U-spin and V-spin, as expected.

Beyond the final states involving two pseudoscalar mesons, the only two-body final states which
are kinematically accessible are $\rho \pi$ and $\omega \pi^0$.
In order to consider these  final states, it necessary to
include vector mesons in the chiral Lagrangian.  For this purpose, we use the
results of~\cite{Terschlusen:2012xw}.
None of the relevant final states can produced directly by a contact interaction involving the
vector meson octet, but
they can be produced through a coupling of dark matter to an intermediate vector meson.
We thus  need the couplings of a vector meson to a vector
spurion, and the trilinear couplings involving at
least one vector meson.
Note that because a neutral vector meson is always produced as an intermediate state, the
quantum numbers of the intermediate vector meson must be the same as those of the initial state.  In particular,
the $\omega$, $\rho^0$ and $\phi$ are only produced as intermediate states if the initial state
is parameterized by the spurion $v^{s1I}$, $v^{tI}$ or $v^{s2I}$, respectively.

The relevant part of the chiral Lagrangian was found in~\cite{Terschlusen:2012xw,Terschlusen:2016kje}, and can be written as
\bea
{\cal L}_{\Phi_{\mu \nu}} &=& -\frac{1}{4} Tr \left[(D^\mu \Phi_{\mu \alpha})
(D_\nu \Phi^{\nu \alpha} ) \right]
+\frac{1}{8} m_V^2 Tr \left[\Phi^{\mu \nu} \Phi_{\mu \nu} \right]
+ \frac{1}{2} f_V Tr \left[\Phi^{\mu \nu} f_{\mu \nu}^+ \right]
+ \frac{i}{2} f_Vh_P
\mathrm{tr}\left( U_\mu \Phi^{\mu\nu}  U_\nu \right)
\nonumber\\
&\,& + \frac{i}{8}h_A\varepsilon^{\mu\nu\alpha\beta}\mathrm{tr}\left(\{\Phi_{\mu\nu},
(D^\tau \Phi_{\tau\alpha}) \} U_\beta \right)
 + \frac{i}{8}h_O\varepsilon^{\mu\nu\alpha\beta}\mathrm{tr}\left([D_\alpha\Phi_{\mu\nu},
 \Phi_{\tau\beta} ] U^\tau \right) ,
\eea
where
\bea
\Phi_{\mu \nu}
&=& \sqrt{2}
 \left(\begin{array}{ccc}
       \frac{\rho_{\mu \nu}}{\sqrt{2}} + \frac{\omega_{\mu \nu}}{\sqrt{2}}
			  & \rho^+_{\mu \nu} & K^{*+}_{\mu \nu} \\
       \rho^-_{\mu \nu} & -\frac{\rho_{\mu \nu}}{\sqrt{2}}
			  + \frac{\omega_{\mu \nu}}{\sqrt{2}} & K^{*0}_{\mu \nu} \\
        K^{*-}_{\mu \nu} & \bar K^{*0}_{\mu \nu} & \phi_{\mu \nu} \\
                \end{array}
              \right) ,
\nonumber\\
f^+_{\mu \nu} &=& \frac{1}{2} \left[e^{\imath \Phi /\sqrt{2}F} \left(
\partial_\mu v_\nu - \partial_\nu v_\mu
\right) e^{-\imath \Phi /\sqrt{2}F}
+ e^{-\imath \Phi /\sqrt{2}F}
\left( \partial_\mu v_\nu  - \partial_\nu v_\mu
\right) e^{\imath \Phi /\sqrt{2}F} \right]         ,
\nonumber\\
U_\mu &=& \frac{1}{2} e^{-\imath \Phi /\sqrt{2}F} \left(\partial_\mu e^{\imath \sqrt{2} \Phi /F} \right) e^{-\imath \Phi /\sqrt{2}F}
-\frac{\imath}{2} e^{-\imath \Phi /\sqrt{2}F} v_\mu   e^{\imath \Phi /\sqrt{2}F}
+\frac{\imath}{2} e^{\imath \Phi /\sqrt{2}F} v_\mu   e^{-\imath \Phi /\sqrt{2}F},
\nonumber\\
D_\alpha \Phi_{\mu \nu} &=& \partial_\alpha \Phi_{\mu \nu} + [\Gamma_\alpha, \Phi_{\mu \nu}],
\nonumber\\
\Gamma_\alpha &=&  \frac{1}{2} e^{-\imath \Phi /\sqrt{2}F} \left(\partial_\mu -\imath v_\mu  \right) e^{\imath \Phi /\sqrt{2}F}
+  \frac{1}{2} e^{\imath \Phi /\sqrt{2}F} \left(\partial_\mu -\imath v_\mu  \right) e^{-\imath \Phi /\sqrt{2}F},
\nonumber\\
f_V &=& (140 \pm 14)~\mev ,
\nonumber\\
m_V &\sim& 0.764~\gev ,
\nonumber\\
h_A &=& 2.33 \pm 0.03 ,
\nonumber\\
h_P &\sim& 1.75 .
\eea

The
terms that produce the couplings of a single vector meson to the vector
spurion
are
\bea \mathcal{L}_{\rm vec}
&=&
- \frac{f_V}{2}  \left( \left( (\partial^\mu v_d^\nu - \partial^\nu v_d^\mu)- (\partial^\mu v_u^\nu - \partial^\nu v_u^\mu)\right) {\rho^0}_{\mu \nu }
- \left( (\partial^\mu v_d^\nu - \partial^\nu v_d^\mu)+ (\partial^\mu v_u^\nu - \partial^\nu v_u^\mu)\right)\omega _{\mu \nu }
-\sqrt{2} (\partial^\mu v_s^\nu - \partial^\nu v_s^\mu) \varphi _{\mu \nu }\right) .
\nonumber\\
\eea

Finally, the cubic couplings between vector and pseudoscalar
meson octets
are
\bea
\Lagr_{\rm int} \supset
 -\frac{\sqrt{2}h_A}{ F}    \varepsilon ^{\mu \nu \alpha \beta }
\bigg{[}{
{1\over 2}\del _{\beta } \pi^ 0}\left( \rho^{0}_{\mu \nu}\del^\tau \omega _{\tau\alpha}
+\omega _{\mu \nu}\del^\tau \rho^{0}_{\tau\alpha}\right)
+{\del _{\beta }  \pi^-}
\left( \rho^{+}_{\mu \nu }\del^\tau\omega _{\tau\alpha}
 +\omega _{\mu \nu}\del^\tau \rho^{+}_{\tau\alpha}\right)
+ c.c
\bigg{]}
\nonumber\\
-\frac{h_O}{\sqrt{2} F}\varepsilon ^{\mu \nu \alpha \beta }
{\bigg{[}} (\del_\alpha {\rho^0}_{\mu \nu })
\bigg{(}{(
\del^\tau\pi^ 0)} \omega _{\tau\beta }
\bigg{)}
+{\rho^0}_{\tau\beta }
\bigg{(}( \del^\tau{\pi^ 0}) \del_\alpha \omega _{\mu \nu }
\bigg{)}
+\left( \del^\tau{\pi^-}
\left(
\omega _{\tau\beta }\del_\alpha {\rho^+}_{\mu \nu }
+
{\rho^+}_{\tau\beta }\del_\alpha \omega _{\mu \nu } \right) + c.c\right)
{
\bigg{]}}
\nonumber\\
+ i\frac{8 {f_V} {h_P}}{F^2} \bigg{[}
2\del _{\mu }{\pi^+} \del _{\nu }{\pi^-} {\rho^0}_{\mu \nu }
+\del _{\mu }{\bar{K}^0} \del _{\nu }{K^0} \left({\rho^0}_{\mu \nu }
-\omega _{\mu \nu }-\sqrt{2}\varphi _{\mu \nu }\right)
-\del _{\mu }{K^-}
\del _{\nu }{K^+} \bigg{(}{\rho^0}_{\mu \nu }+\omega _{\mu \nu }-\sqrt{2}\varphi _{\mu \nu }\bigg{)}
\bigg{]} .
\eea
We have suppressed terms with no neutral vector meson.  Interestingly, none of the relevant matrix
elements depend on the parameter $h_O$.  One can easily see from the Feynman rules in Appendix~\ref{crosssectionappendix} that,
if the center-of-mass momentum of the initial state is purely time-like, then
the only non-vanishing polarizations of the intermediate vector meson are mixed time-space.  Similarly, in the frame of reference
where the outgoing vector meson is at rest, its polarization tensor is also necessarily mixed time-space.  It is then clear that
all of the terms proportional to $h_O$ vanish.

The  rates for dark matter to annihilate to primary mesons can now be calculated
straightforwardly from this Lagrangian; the details are given in Appendix
\ref{crosssectionappendix}.
These primary mesons can then decay through multiple decay modes to produce photons.
 These
can be multistep decay processes; for example, the kaon can decay to pions which
subsequently decay to photons.
In our case, the primary mesons are $\pi^0, \pi^{\pm}, K^0, \bar K^0, K^{\pm}, \rho, \omega$, and
we need to find the photon spectra produced in their decays.
The $\pi^0$ decays to two photons essentially 100\% of the time, and
the $\pi^\pm$ essentially never produce photons.
The $\rho^0$ decays primarily to charged pions, and hence  does not produce photons.
But decays of $\rho^\pm$ typically produce $\pi^0$ as well as $\pi^\pm$, with the
subsequent the decays of $\pi^0$ yielding photons.
For the kaons and the omega, we use the decay modes
 found in the Particle Data Book~\cite{pdg}. We
have tabulated in Table~\ref{Tab:BranchingFraction} the important decay modes of
the mesons which we have considered along with their branching ratios. Decay modes which are not
expected to produce a significant number
of photons (e.g. decays involving only charged pions) are not shown.

\begin{table}[h!]
  \begin{center}
    \begin{tabular}{|c|c|c|}
      \hline
      \multirow{4}{*}{$K^+$}
      & $\pi^0e^+\nu$ & 5\%\\
		  & $\pi^0\mu^+\nu$ & 3.4\%\\
			& $\pi^+\pi^0$& 20.7\%\\
			& $\pi^+\pi^0\pi^0$& 1.7\%\\
      \hline
      \multirow{1}{*}{$\rho^\pm$}
			& $\pi^\pm \pi^0$& 100\%\\
      \hline
    \end{tabular}
		    \begin{tabular}{|c|c|c|}
      \hline
      \multirow{1}{*}{$K^S$}
			& $\pi^0\pi^0$& 30.7\%\\
      \hline
		\multirow{2}{*}{$K^L$}
			& $\pi^0\pi^0\pi^0$& 19.5\%\\
			& $\pi^+\pi^-\pi^0$&  12.5\%\\
      \hline
      \multirow{2}{*}{$\omega$}
			& $\pi^+\pi^-\pi^0$& 89\%\\
			& $\pi^{0}\gamma$& 8\%\\
      \hline
    \end{tabular}
		    \caption{The relevant decay modes and branching fractions for the mesons produced from
dark matter annihilation/decay through the quark vector current portal.}
\label{Tab:BranchingFraction}
  \end{center}
\end{table}

For the two-body decays, the decay spectrum at rest is fixed by kinematics.
The three-body decays are described by a Dalitz plot, which parametrizes the
decay kinematics in terms of the final state energies. We use these to
find the decay spectrum at rest for each decay mode of each meson. These details are
presented in  Appendix \ref{decayspectrumappendix}.
Finally,
these mesons are produced and decay
 with large boosts which also modify the spectrum. The boosting procedure is
general, and is described in Appendix \ref{boostdecayspectrumappendix}.

\begin{figure}[t]
\centering
\includegraphics[scale=0.5]{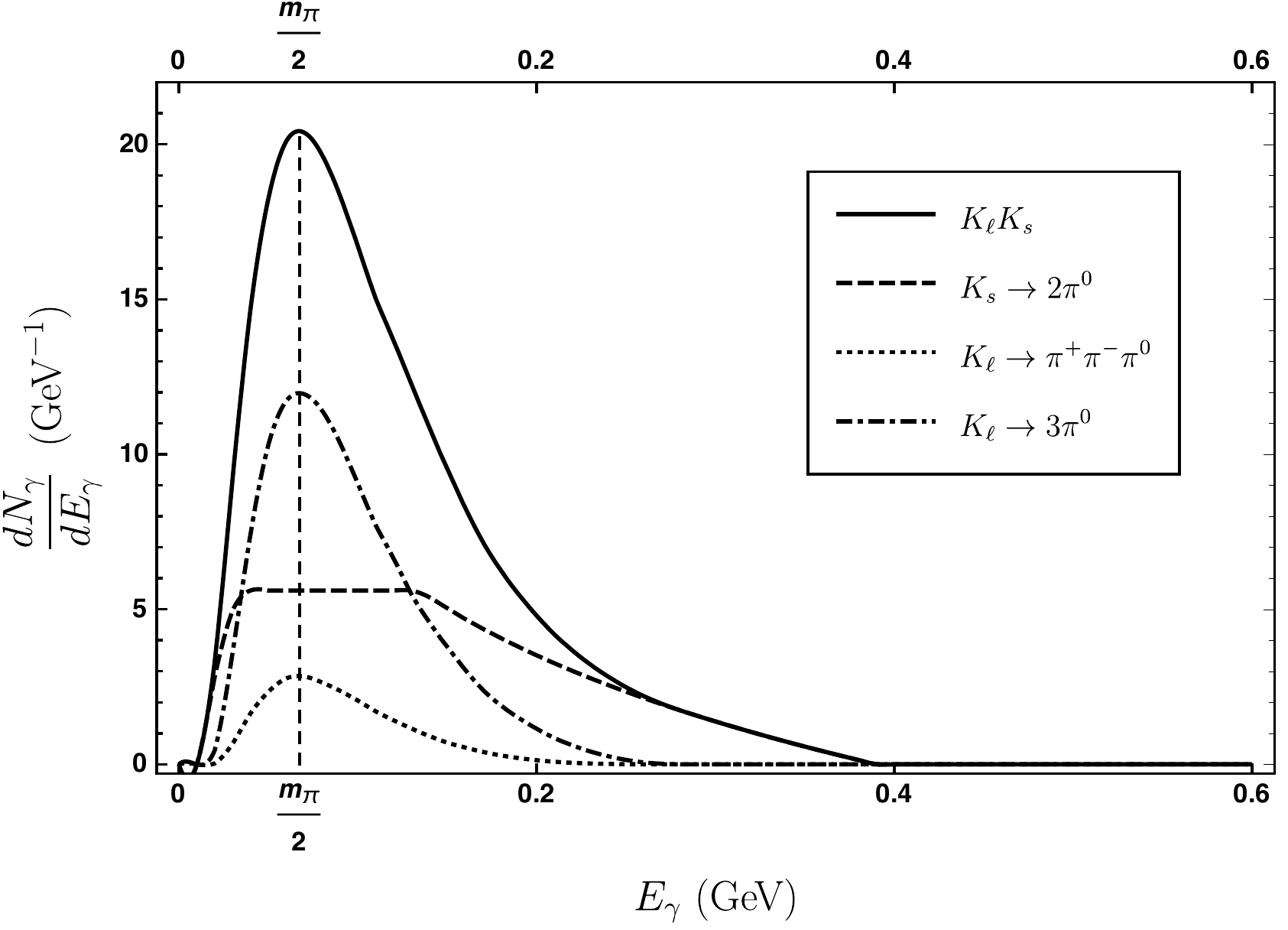}
\caption{Photon spectrum from neutral kaon production ($K_L K_S$).
The center-of-mass energy has been taken to be $1.14$ GeV.
\label{Fig:KLKS_Spectrum}
}
\end{figure}

For illustrative purposes, we show in Figure~\ref{Fig:KLKS_Spectrum} the photon spectrum
obtained through the production of neutral kaons produced from dark matter annihilation or decay, assuming the
$E_{cm} =1.14$ GeV. The individual spectra have been
weighted according to the branching fractions.  It is interesting to compare this spectrum to
those found in~\cite{Kumar:2018heq}, for the case where dark matter coupled to scalar, pseudoscalar
or axial-vector quark currents.  In those cases, even though the center-of-mass energy was taken
to be $\leq 1~\gev$, the typical photon energy was nearly 4 times larger than found here.  The reason
is because in the case of a scalar, pseudoscalar or axial-vector spurion, one can produce an
$\eta$ in the final state, whose decays yield photons.  In the case of a vector spurion, however, almost
all photons arise from $\pi^0$ decay.  As is discussed in Appendix D, the photons produced by $\pi^0$ decay
yield a typical photon energy which is considerably smaller than that of the photons
produced from $\eta$ decay.

\section{Comparisons to Observations}

 Following~\cite{Kumar:2018heq}, we consider constraints on the obtained photon spectrum
 from observations of diffuse photon emission, and from future observations of
photon emission from a dwarf galaxy (Draco).
We will take as a benchmark, an experiment with a fractional $1 \sigma$ energy resolution of
$\epsilon = 0.3$
	and an exposure of $3000~\cm^2~\yr$.
We also assume that the experimental
angular resolution is smaller than the size of Draco ($1.3^\circ$).

For diffuse emission, we restrict attention
to latitudes greater than $> 20^\circ$. In this region, and in the energy range
0.8 MeV - 1 GeV, the isotropic flux observed by COMPTEL and EGRET can be well
fit~\cite{Boddy:2015efa,Strong:2004de} to
the function
\bea
\frac{d^2 \Phi^{iso.}}{d\Omega~dE_{obs.}} &=& 2.74 \times 10^{-3}
\left( \frac{E_{obs.}}{\mev} \right)^{-2.0} \cm^{-2} \s^{-1} \sr^{-1} \mev^{-1}.
\label{eq:ObservedDiffuseFlux}
\eea

The expected number of observed events between the energies $E_-$ and $E_+$ is then given by
\bea
N_{O} &=& 8.6 \times 10^{4} \left(\frac{\mev}{E_-} - \frac{\mev}{E_+} \right)
\frac{(I_{exp.} \Delta \Omega)}{\cm^2~\yr~\sr} .
\eea

To estimate the constraints on the model from diffuse emission, we consider energy bins
set by the resolution through $E_+ - E_- = \epsilon(E_+ + E_-)$. We impose
the condition that within any such energy bin, the number of
expected signal events should not exceed the number of
expected observed events i.e., we require $N_S < N_O$.  This constitutes a
conservative bound; since we assume no knowledge of the background photon spectrum,
we conservatively assume that the background could be negligible in any energy bin.
While in principle we can scan over all possible energy bins,
we note that almost all photons arise from tertiary processes in which the decay of a heavier meson
produces a $\pi^0$, whose decay in turn produces a pair of photons.\footnote{Note, there will also
be a small number of
secondary photons produced through the process $\omega \rightarrow \pi^0 \gamma$.  But since the
branching fraction
for this process is small, and the resulting feature is in any case not sharp, there is little
reason to consider it further.}
As discussed in Appendix D,
a signal of this form has a peak
at $m_{\pi} / 2$, so we shall
center our bin at $E_0 = m_\pi/2$.

For a dwarf spheroidal galaxy, such as Draco, the background consists of all photons
emitted by astrophysical processes as well as dark matter annihilation/decay along the
line of sight to the dwarf, but not within the dwarf.  This background can be estimated purely
from data by considering the observed flux in the direction of the dSph, but slightly
off-axis~\cite{GeringerSameth:2011iw,Mazziotta:2012ux,GeringerSameth:2014qqa,Boddy:2018qur,Albert:2017vtb}).
Although one would follow this approach in an actual analysis of data from a future instrument,
for our purposes, we can estimate the background flux to be roughly the same as observed
diffuse flux given in eq.~\ref{eq:ObservedDiffuseFlux}.
If we assume that the number
of observed events is the same as the expected number of background events, then
a model can be ruled out at $n-\sigma$ confidence level if $N_S > n \sqrt{N_O}$.

\section{Results}

The differential photon flux from dark matter annihilation or decay is
\bea
\frac{d^2 \Phi}{d\Omega~dE_\gamma} &=& \frac{\Xi^{ann.,dec.}}{4\pi m_X}
{\bar J}^{ann.,dec.}\frac{dN_\gamma}{dE_\gamma} ,
\eea
where
\bea
\Xi^{ann.} &=& \frac{\langle \sigma_A  v \rangle}{2m_X} ,
\nonumber\\
\Xi^{dec.} &=& \Gamma ,
\eea
and  where $\bar J^{ann.,dec.}$ is the average $J$-factor of the target for either dark matter
annihilation or dark matter decay.

Here, for diffuse emission, the average $J$-factor is~\cite{Cirelli:2010xx}
\bea
\bar J_{dif.}^{ann.} &=& 3.5 \times 10^{21}~\gev^2 ~\cm^{-5} \sr^{-1} ,
\nonumber\\
\bar J_{dif.}^{dec.} &=& 1.5 \times 10^{22}~\gev ~\cm^{-2} \sr^{-1} ,
\label{eq:JFactorDiffuse}
\eea
while for Draco,  the average $J$-factor is~\cite{Geringer-Sameth:2014yza}
\bea
\bar J_{Draco}^{ann.} &=& 6.94 \times 10^{21} ~\gev^2 ~\cm^{-5} ~\sr^{-1} ,
\nonumber\\
\bar J_{Draco}^{dec.} &=& 5.77 \times 10^{21} ~\gev ~\cm^{-2} ~\sr^{-1} .
\label{eq:JFactorDraco}
\eea

We account for instrumental energy resolution by convolving the injected photon spectrum
with a Gaussian smearing function
\bea
R_\epsilon (E_{obs.},E_\gamma) &=& \frac{1}{\sqrt{2\pi} \epsilon E_\gamma}
\exp \left(-\frac{(E_{obs.}-E_\gamma)^2}{2\epsilon^2 E_\gamma^2} \right) .
\eea

In terms of the
  the exposure $I_{exp.}$, and  the solid angle $\Delta \Omega$,
the number of events expected  within the energy window $E_- \leq E_{obs.} \leq E_+$ is
\bea
N_{S} &=& \frac{\Xi^{ann.,dec.}}{4\pi m_X} {\bar J}^{ann.,dec.} (I_{exp.}
\Delta \Omega)  \int_{E_-}^{E_+} dE_{obs.} \int_0^\infty
dE_\gamma ~ \frac{dN_\gamma}{dE_\gamma} R_\epsilon (E_{obs.},E_\gamma)
\eea

We now consider the two models described by equations (\ref{eq:IntLagrangian}) and (\ref{eq:IntLagrangian2}). In each case,
we find the photon spectrum from decay/
annihilations and impose bounds from observations. We consider two classes of models.
In the first class we assume isospin is a good symmetry, and we  set $\alpha_u=\alpha_d$.
In the second class we set $\alpha_s=0$.

\begin{figure}[t]
\centering
\includegraphics[height=5cm, width=5cm 
]{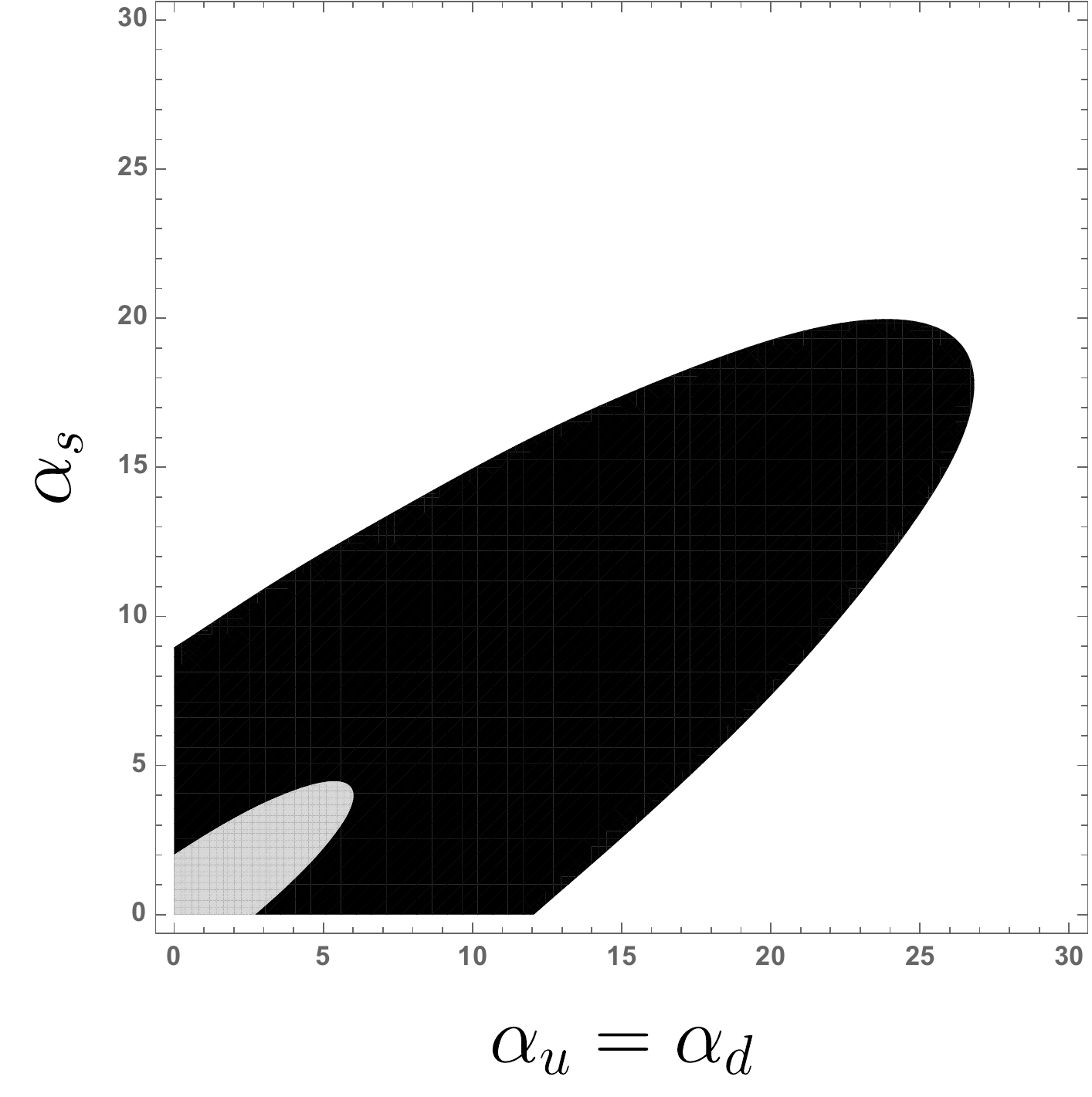}
\qquad\qquad
\includegraphics[height=5cm, width=5cm 
]{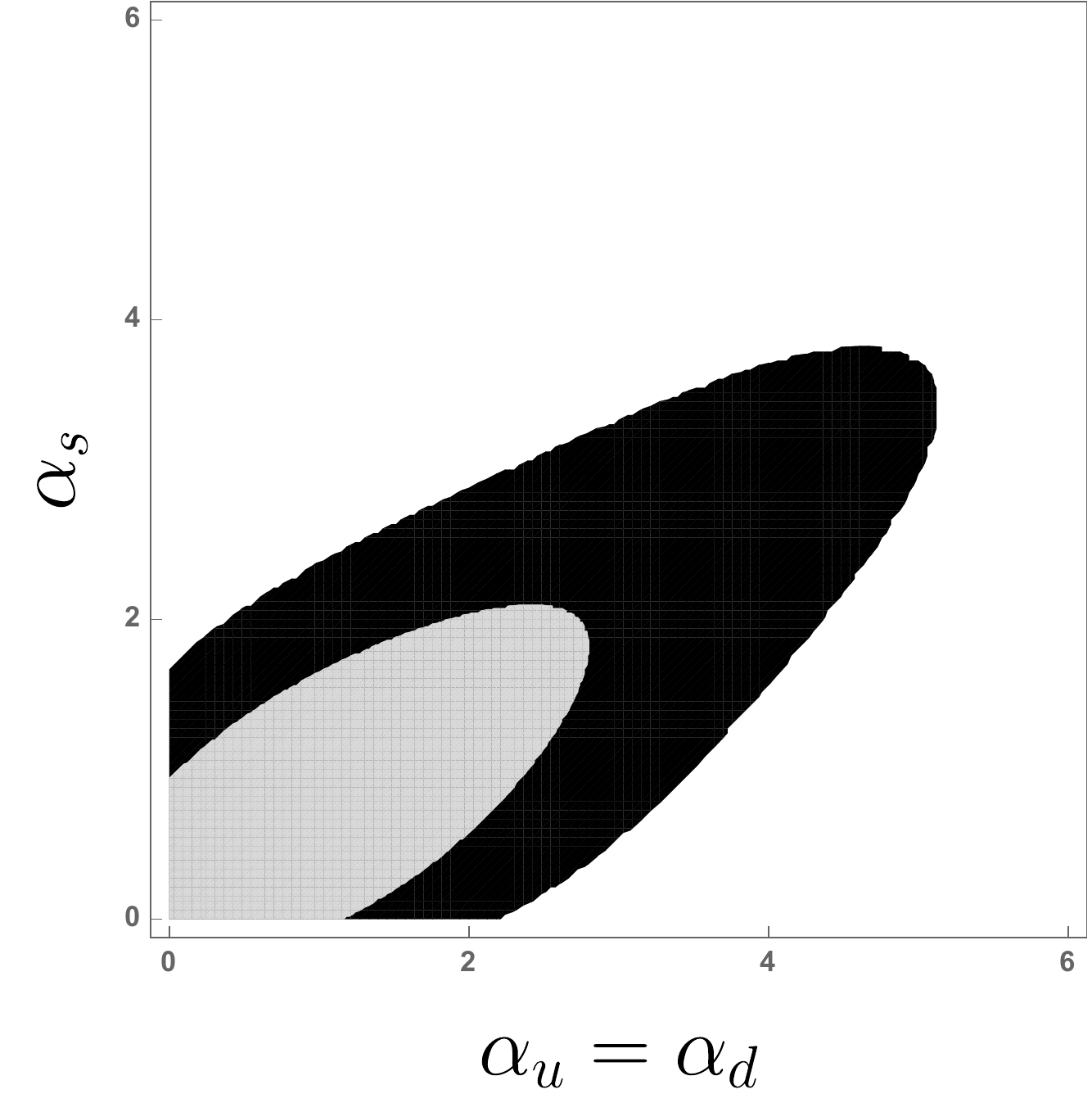}
\caption{Bounds on the parameter space from annihilation (left panel, $\Lambda = 100~\gev$) and decay (right panel, $g=10^{-24}$)
on theories with $\alpha_u=\alpha_d$, assuming $E_{cm} = 1.14~\gev$.
The region in black is allowed by constraints on diffuse emission, while the grey region demarcates
the sensitivity of a search for emission from Draco.
}
\label{resultsfig1}
\end{figure}

\begin{figure}[t]
\centering
\includegraphics[height=4cm, width=5cm 
]{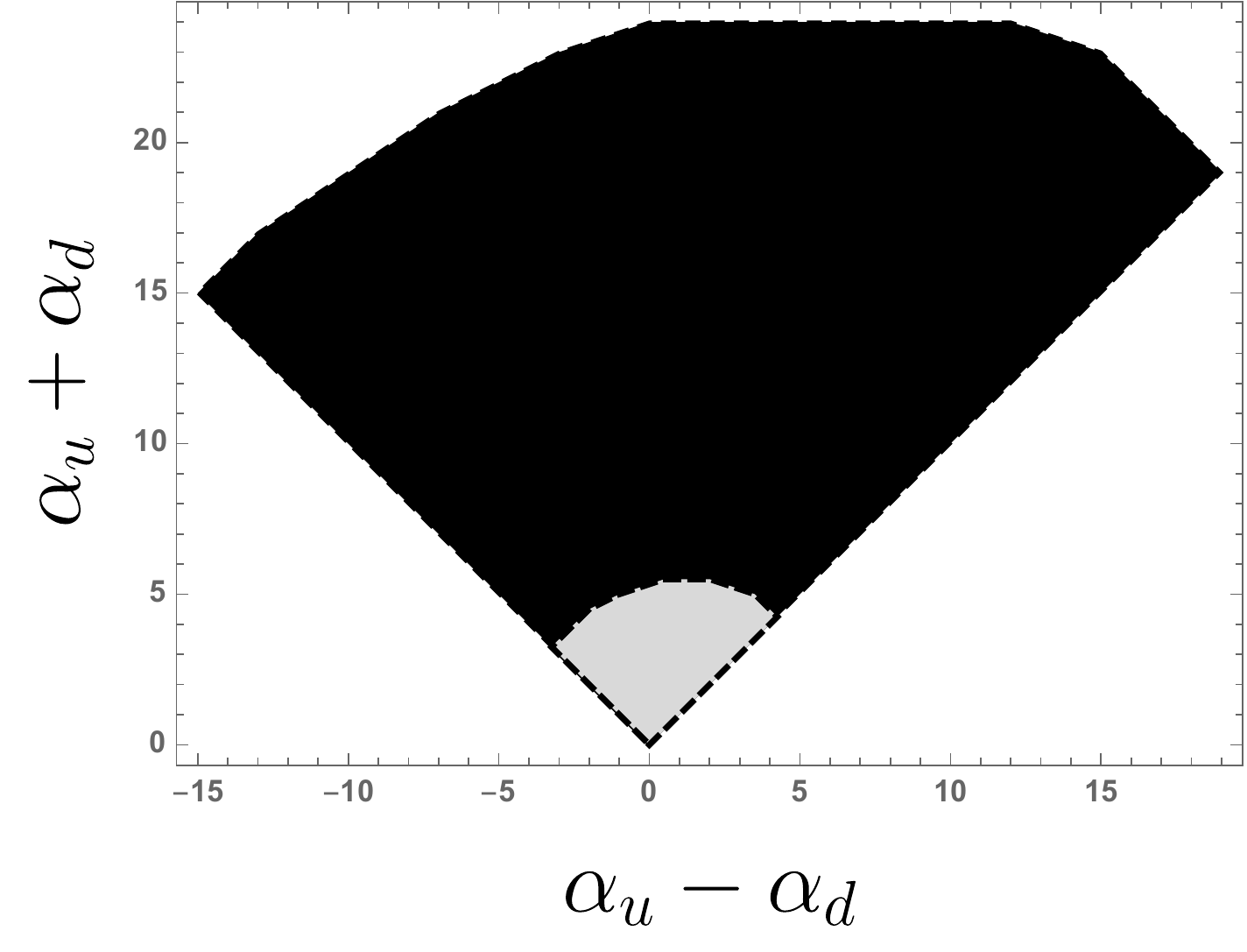}
\qquad\qquad
\includegraphics[height=4cm, width=5cm 
]{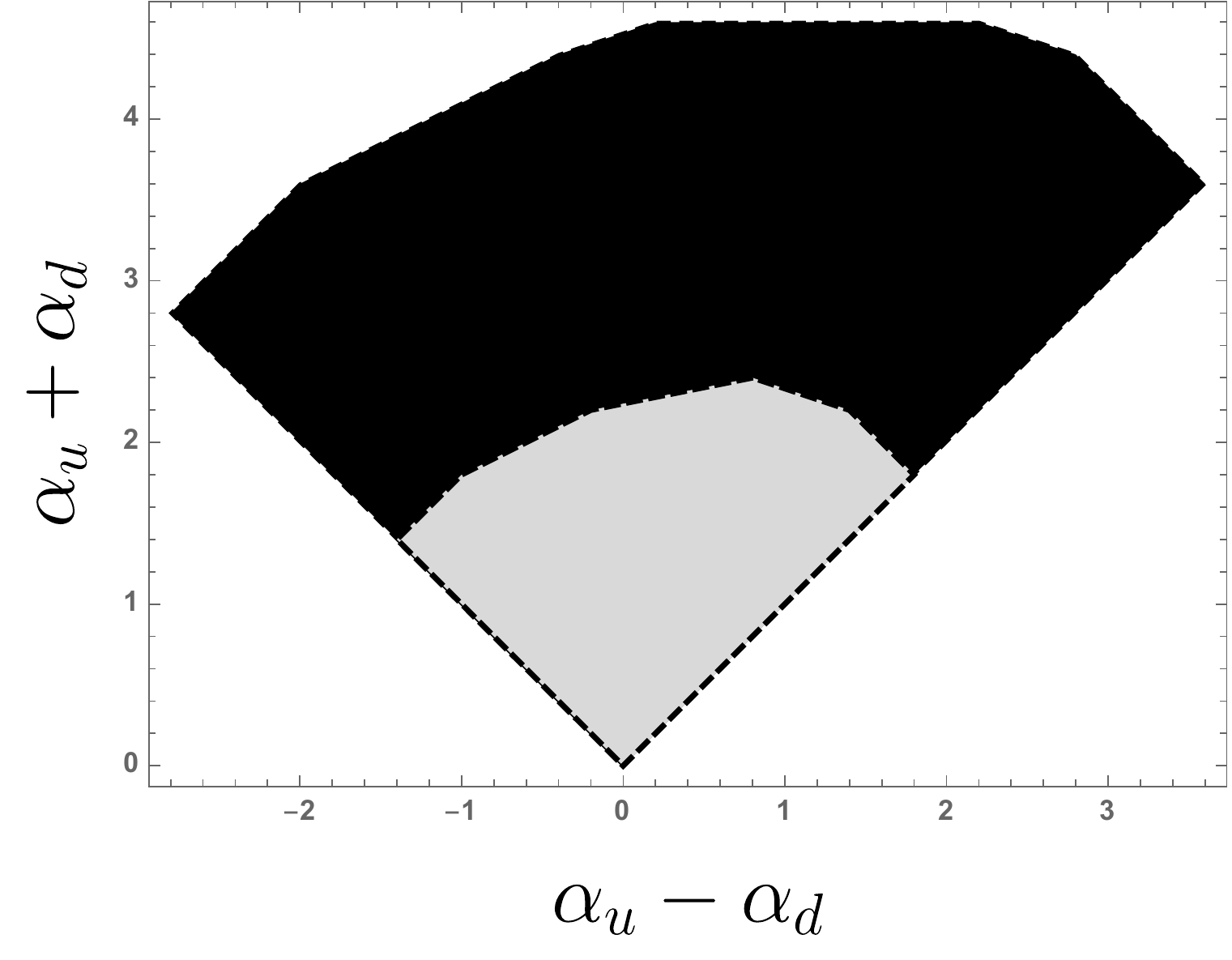}
\caption{Bounds on the parameter space from annihilation (left panel, $\Lambda = 100~\gev$) and decay (right panel, $g=10^{-24}$)
on theories with $\alpha_s =0$, assuming $E_{cm} = 1.14~\gev$.
The region in black is allowed by constraints on diffuse emission, while the grey region demarcates
the sensitivity of a search for emission from Draco.
}
\label{resultsfig2}
\end{figure}

Bounds on the parameter space of the first class of models
are shown in Figure~\ref{resultsfig1}. Bounds on the parameter space of the
second class of models
are shown in Figure~\ref{resultsfig2}.
In each case, we have normalized the couplings to have
$g=10^{-24}$ in equation (\ref{eq:IntLagrangian}), and $\Lambda=100~\gev$ in equation (\ref{eq:IntLagrangian2}).
In all cases we have set the center-of-mass energy to be $E_{cm} = 1.14~\gev$.
Because the energy range between the $\rho \pi $ threshold ($\sim 0.91~\gev$) and our
upper cutoff ($1.15~\gev$) is relatively narrow, these constraints do not change dramatically
as we change the center-of-mass energy.  But there is one feature worthy of note; if
$E_{cm} \sim 1.02~\gev$ and if dark matter couples to a strange quark vector current,
then dark matter can annihilate through an intermediate $\phi$ near resonance.  In this case,
the constraints on $\alpha_s$ which could be obtained from indirect detection would improve
dramatically.

Note that constraints from Draco are tighter than those from diffuse emission, but more so in the
case of dark matter annihilation than in the case of decay.  This is what one would expect, as the
annihilation rate scales as the square of the density, whereas the decay rate scales only with the density.

Note also that, if the dark matter coupling does not break isospin, then the constraints on the couplings
are weakest if $\alpha_u = \alpha_d =\alpha_s$.  This can easily be understood from equation (\ref{eq:PSMesonCoupling});
if all light quark couplings are identical, the dark matter coupling to the pseudoscalar meson
octet vanishes.  In this limit, the only remaining channels which produce photons are
$\omega \pi^0$ and $\rho \pi$.  The constraints on dark matter decay (annihilation) obtained from diffuse emission
in this limit correspond to $\tau \sim {\cal O}(10^{25})~\s$ ($\langle \sigma_A v \rangle
\sim {\cal O}(10^{-25}) ~\cm^3 /\s$).
A similar sensitivity to dark matter decay (annihilation) in Draco could yield bounds of
$\tau \sim {\cal O}(10^{26})~\s$ ($\langle \sigma_A v \rangle
\sim {\cal O}(10^{-26}) ~\cm^3 /\s$).
The sensitivity to dark matter decay thus exceeds that obtainable by
Planck~\cite{Aghanim:2018eyx} (from the effect of late-time energy injection on the CMB) by about an order of magnitude.
But bounds from Planck are much tighter in the case of dark matter annihilation, and exceed the sensitivity we
would obtain from a search for emission from Draco by about two orders of magnitude.
But if more nearby dwarf spheroidal galaxies are
discovered, then one might hope that a stacked analysis of all dwarfs could provide competitive sensitivity.

For the isotropic flux, we have used the best fit to data from COMPTEL and EGRET.  But future instruments,
with larger exposure and better angular resolution, may be able to identify more point sources.  In this case,
masking these point sources could result in a significant reduction of the remaining isotropic flux, and the
sensitivity of this analysis would improve accordingly.

Finally, we note that direct detection experiments and LHC mono-anything searches are completely unconstraining
for models of dark matter decay.  As noted in~\cite{Kumar:2018heq}, LHC searches are also do not constrain models
of dark matter annihilation if the mass scale of the particle  mediating the DM-quark interaction is
${\cal O}(\gev)$.  But for  $m_\chi \sim 500~\mev$, the upper bound on spin-independent (SI) dark matter-nucleon
scattering found by CRESST~\cite{Kluck:2017hnn}
is $\sigma_{\rm SI} \sim {\cal O}(10^{-1})~\pb$.  For a scenario of dark matter annihilation
through coupling to the quark vector portal, DM-nucleon scattering would be spin and velocity-independent.  If
we set $\alpha_u=\alpha_d \sim 3$, $\alpha_s=0$ (the limit of sensitivity for a search for emission from Draco), one
would find $\sigma_{\rm SI} \sim {\cal O}(10)~\pb$, about two orders of magnitude above current bounds from CRESST.  However,
CRESST's sensitivity is greatly reduced for the case of isospin-violating dark matter~\cite{Chang:2010yk,Feng:2011vu,Feng:2013fyw}
($\alpha_u=-\alpha_d$, $\alpha_s=0$),
and is essentially unconstraining for the parameter space of interest here.  Similarly, if dark matter only couples to the
strange quark vector current ($\alpha_u = \alpha_d =0$), then the SI scattering cross section vanishes, and there are no
meaningful constraints from direct detection.

\section{Conclusions}

In conclusion, we have considered dark matter coupled to Standard Model quarks through
vector currents of the form $\bar{q}\gamma^\mu q$. We have utilized the chiral Lagrangian to
obtain couplings of the dark matter to mesons, and found the photon spectrum from
dark matter decay (if the dark matter itself is a vector) or
from annihilation (where it is a Dirac fermion).

We have found that current observations of the diffuse photon spectrum already can be used to
constrain the parameter  space of these models.
Future observations of dwarf spheroidal galaxies will significantly improve these bounds or
may discover these models.
In the case of dark matter decay, these constraints
exceed those obtainable from Planck by about an order of magnitude.

There are several ways to extend this analysis.  We have considered final states with at most
two mesons, and have worked to ${\cal O}(p^2)$ in the chiral Lagrangian.  Consideration of  4-meson
final states, and higher order terms in the chiral Lagrangian momentum expansion, would yield subleading
corrections to the photon spectrum which may nevertheless be significant, espeically below the $\rho \pi$
threshold.
Similarly, if we consider higher center-of-mass energies, then new final state particles will be
accessible, including the $\phi$ and glueball states.
We hope to return to this in future work.

We have found that if dark matter couples to a vector quark current, then the typical photon
energy is roughly a factor of 4 smaller than in the scenarios considered in~\cite{Kumar:2018heq},
wherein dark matter coupled to scalar, pseudoscalar or axial-vector currents.  This result is largely
independent of the center-of-mass energy of the process, but is instead dictated by the Lorentz and
flavor structure of the dark matter-quark current coupling, which determines whether or not an $\eta$
can be produced in the final state.  It would be interesting to study more quantitatively the ability of
future experiments to definitively determine the Lorentz and flavor structure of the DM-quark coupling
from the photon distribution.

Finally, we note that, although we have assumed an energy resolution of $30\%$ and an exposure
of $3000~\cm^2~\yr$, upcoming experiments may well exceed these performance benchmarks.  Energy
resolutions of as low as $10\%$ may be possible~\cite{Buckley,Caputo}, as well as exposures which are up to an order of
magnitude larger~\cite{Buckley}.

\section{Acknowledgements}

We are grateful to James H.~Buckley, Regina Caputo and Xerxes Tata for useful discussions.
J.~K. is supported in part by DOE grant DE-SC0010504.
A.~R. is supported in part by NSF  Grant PHY-1620638.


\appendix

\section{Symmetry considerations}
\label{symmetries}

Since we assume that the primary dark matter annihilation/decay process does
not involve weak interactions, we will
find that the hadronic final state has vanishing strangeness, and has the same
parity, charge conjugation and angular
momentum quantum numbers as the initial state.
We also, for simplicity, focus only on  primary
annihilation/decay processes which produce at most
two mesons.  As such, the only kinematically accessible neutral
two-body final states with vanishing net strangeness are
$\pi \pi$, $\eta \eta$, $\eta \pi^0$, $\rho \pi$, $\omega \pi^0$,
$K^+ K^-$, $K^0 \bar K^0$.

The quantum numbers of the Standard Model final state can be determined in general
from the quantum numbers of the dark matter initial state (for example, see~\cite{Kumar:2013iva}).
If dark matter couples to a vector quark current, the final state necessarily has
$J^{PC} = 1^{--}$.  As such, the
$\pi^0 \pi^0$, $\eta \pi^0$, $\eta \eta$, $K_L K_L$ and $K_S K_S$ final states are forbidden.
 As a result, the only final states
which we need consider are
$\pi^+ \pi^-$, $K^+ K^-$, $K^L K^S$, $\rho \pi$, and $\omega \pi^0$.  Since all of the mesons in question
have odd intrinsic parity, each state transforms under parity as $(-1)^L$, and must thus have
orbital angular
momentum $L=odd$.  From the symmetry of the wavefunction,
we then see that the $\pi^+ \pi^-$ state must have isospin $I=1$, $I_3 =0$.
Similarly, we can classify the two-kaon final states in terms of their
isospin quantum numbers:
\bea
(K K)_s &=& \frac{1}{\sqrt{2}} \left(|K^+ K^- \rangle + |K_L K_S \rangle \right) \qquad I=0,
\nonumber\\
(K K)_t &=& \frac{1}{\sqrt{2}} \left(|K^+ K^- \rangle - |K_L K_S \rangle \right) \qquad I=1,~ I_3=0 .
\eea
Note, the choice of which relative sign corresponds to the singlet or triplet state is a convention,
which depends on the normalizations of the $|K^+ K^- \rangle$ and $|K_L K_S \rangle$ states.
The $\omega \pi^0$ state is necessarily $I=1$, $I_3=0$, while
the $\rho \pi$ state is $I=0,1$ or 2, $I_3=0$.
But since the isospin quantum numbers of the final state should be the same as those of
the quark current to which the vector spurion couples, we should only be only be able to
produce $\rho \pi$ states with $I=0$  or 1; the $I=2$ $\rho \pi$ state
should be inaccessible.

Thus, we essentially have six final states, two of which have
$I=I_3=0$ ($(KK)_s$ and $(\rho \pi)_s$), and four of which have $I=1$, $I_3=0$
($(\pi \pi)_t$, $(K K)_t$, $(\rho \pi)_t$ and $(\omega \pi)_t$).  Note that the $(\pi \pi)_t$
state produces no photons, while the $(K K)_s$ and $(K K)_t$ states produce identical photon spectra.

Isospin is an $SU(2)$
subgroup
of the QCD flavor symmetries which relates $u$- and $d$-quarks.  But similarly, there are
$SU(2)$ subgroups
which related $d$- and $s$-quarks (U-spin) and $u$- and $s$-quarks (V-spin).  We thus find that the
$K^+ K^-$ state must have V-spin $I^V =1$, $I^V_3=0$, while the states
$(1/\sqrt{2})(|\pi^+ \pi^-\rangle \pm |K_L K_S\rangle)$  have $I^V=I_3^V=0$ and
$I^V=1$, $I^V_3=0$, respectively.
Similarly, the $K^L K^S$ state must have U-spin $I^U =1$, $I^U_3=0$, while the
states $(1/\sqrt{2})(|\pi^+ \pi^-\rangle \pm |K^+ K^-\rangle)$  have $I^U=I_3^U=0$ and
$I^U=1$, $I^U_3=0$, respectively.  Note that U-spin and V-spin are not useful
for classifying the $\rho \pi$ and $\omega \pi^0$ final states, since the $\rho$ and
$\omega$ transform into kinematically inaccessible states under U-spin and V-spin; essentially,
the $\rho$ and $\omega$ are necessarily close enough to threshold that the strange quark mass
cannot be taken as negligible.

\section{Feynman Rules}
\label{crosssectionappendix}

We here collect the relevant Feynman rules. For all rules below, all momenta on the left are entering the vertex, and those on the right are exiting.
\vskip -0.1 cm
\bea
\begin{tikzpicture}
		\draw[thin] (0,0)  -- (0.5,0.5);
		\draw[thin,<-]  (0.5,0.5) -- (1,1);
				\draw[thin,->] (0,2)  -- (0.5,1.5);
		\draw[thin]  (0.5,1.5) -- (1,1);
	\draw[dashed] (2,2)  -- (1.5,1.5);
	\draw[dashed] (1.5,1.5)  -- (1,1);
						\draw[dashed] (2,0)  -- (1.5,0.5);
			\draw[dashed] (1.5,0.5)  -- (1,1);
	\node at (2.2,1.7) {$K^L,k$};
	\node at (2.2,0.5) {${K}^S, k'$};
				\node at (-0.5,1.9) {$\chi_\beta,p$};
	\node at (-0.5,0.1) {$\bar{\chi}^\alpha,p'$};
\end{tikzpicture}
\begin{array}{c}
 \frac{1}{\Lambda^2}(\gamma^\mu)_\alpha ^{~\beta} (\alpha_d - \alpha_s)(k - k')_\mu
\\ \
\\ \
\\ \
\\ \
\end{array}
\quad \quad
\begin{tikzpicture}
		\draw[thin] (0,0)  -- (0.5,0.5);
		\draw[thin,<-]  (0.5,0.5) -- (1,1);
				\draw[thin,->] (0,2)  -- (0.5,1.5);
		\draw[thin]  (0.5,1.5) -- (1,1);
	\draw[dashed] (2,2)  -- (1.5,1.5);
	\draw[dashed,->] (1,1)  -- (1.5,1.5);
			\draw[dashed,->] (2,0)  -- (1.5,0.5);
			\draw[dashed] (1.5,0.5)  -- (1,1);
	\node at (2.2,1.7) {$K^+,k$};
	\node at (2.2,0.5) {${K}^-, k'$};
				\node at (-0.5,1.9) {$\chi_\beta,p$};
	\node at (-0.5,0.1) {$\bar{\chi}^\alpha,p'$};
\end{tikzpicture}
\begin{array}{c}
 \frac{1}{\Lambda^2}(\gamma^\mu)_\alpha ^{~\beta} (\alpha_d - \alpha_s)(k - k')_\mu
\\ \
\\ \
\\ \
\\ \
\end{array}
\eea

\bea
\begin{tikzpicture}
		\draw[dashed] (0,0)  -- (0.5,0.5);
		\draw[dashed]  (0.5,0.5) -- (1,1);
				\draw[dashed] (0,2)  -- (0.5,1.5);
		\draw[dashed]  (0.5,1.5) -- (1,1);
			\node at (-0.5,1.9) {$K^L,k$};
	\node at (-0.5,0.1) {${K}^S,k'$};
		\node at (2,0.7) {$\rho_{\mu\nu}$};
					\draw[dashed] (1,1)  -- (2,1);
\end{tikzpicture}
\begin{array}{c}
i\frac{4{f_V} {h_P}}{F^2}(k^\mu (k')^\nu - k^\nu (k')^\mu)
\\ \
\\ \
\\ \
\\ \
\end{array}
\quad \quad
\begin{tikzpicture}
		\draw[dashed,->] (0,0)  -- (0.5,0.5);
		\draw[dashed]  (0.5,0.5) -- (1,1);
				\draw[dashed] (0,2)  -- (0.5,1.5);
		\draw[dashed,<-]  (0.5,1.5) -- (1,1);
			\node at (-0.5,1.9) {$K^+,k$};
	\node at (-0.5,0.1) {${K}^-,k'$};
		\node at (2,0.7) {$\rho_{\mu\nu}$};
					\draw[dashed] (1,1)  -- (2,1);
\end{tikzpicture}
\begin{array}{c}
i\frac{4{f_V} {h_P}}{F^2}(k^\mu (k')^\nu - k^\nu (k')^\mu)
\\ \
\\ \
\\ \
\\ \
\end{array}
\eea

\bea
\begin{tikzpicture}
		\draw[dashed] (0,0)  -- (0.5,0.5);
		\draw[dashed]  (0.5,0.5) -- (1,1);
				\draw[dashed] (0,2)  -- (0.5,1.5);
		\draw[dashed]  (0.5,1.5) -- (1,1);
			\node at (-0.5,1.9) {$K^L,k$};
	\node at (-0.5,0.1) {${K}^S,k'$};
		\node at (2,0.7) {$\omega_{\mu\nu}$};
					\draw[dashed] (1,1)  -- (2,1);
\end{tikzpicture}
\begin{array}{c}
-i\frac{4{f_V} {h_P}}{F^2}(k^\mu (k')^\nu - k^\nu (k')^\mu)
\\ \
\\ \
\\ \
\\ \
\end{array}
\quad \quad
\begin{tikzpicture}
		\draw[dashed,->] (0,0)  -- (0.5,0.5);
		\draw[dashed]  (0.5,0.5) -- (1,1);
				\draw[dashed] (0,2)  -- (0.5,1.5);
		\draw[dashed,<-]  (0.5,1.5) -- (1,1);
			\node at (-0.5,1.9) {$K^+,k$};
	\node at (-0.5,0.1) {${K}^-,k'$};
		\node at (2,0.7) {$\omega_{\mu\nu}$};
					\draw[dashed] (1,1)  -- (2,1);
\end{tikzpicture}
\begin{array}{c}
-i\frac{4{f_V} {h_P}}{F^2}(k^\mu (k')^\nu - k^\nu (k')^\mu)
\\ \
\\ \
\\ \
\\ \
\end{array}\eea

\bea
\begin{tikzpicture}
		\draw[dashed] (0,0)  -- (0.5,0.5);
		\draw[dashed]  (0.5,0.5) -- (1,1);
				\draw[dashed] (0,2)  -- (0.5,1.5);
		\draw[dashed]  (0.5,1.5) -- (1,1);
			\node at (-0.5,1.9) {$K^L,k$};
	\node at (-0.5,0.1) {${K}^S,k'$};
		\node at (2,0.7) {$\phi_{\mu\nu}$};
					\draw[dashed] (1,1)  -- (2,1);
\end{tikzpicture}
\begin{array}{c}
i\sqrt{2}\frac{4{f_V} {h_P}}{F^2}(k^\mu (k')^\nu - k^\nu (k')^\mu)
\\ \
\\ \
\\ \
\\ \
\end{array}
\quad \quad
\begin{tikzpicture}
		\draw[dashed,->] (0,0)  -- (0.5,0.5);
		\draw[dashed]  (0.5,0.5) -- (1,1);
				\draw[dashed] (0,2)  -- (0.5,1.5);
		\draw[dashed,<-]  (0.5,1.5) -- (1,1);
			\node at (-0.5,1.9) {$K^+,k$};
	\node at (-0.5,0.1) {${K}^-,k'$};
		\node at (2,0.7) {$\phi_{\mu\nu}$};
					\draw[dashed] (1,1)  -- (2,1);
\end{tikzpicture}
\begin{array}{c}
i\sqrt{2}\frac{4{f_V} {h_P}}{F^2}(k^\mu (k')^\nu - k^\nu (k')^\mu)
\\ \
\\ \
\\ \
\\ \
\end{array}\eea

\bea
\begin{tikzpicture}
		\draw[thin] (0,0)  -- (0.5,0.5);
		\draw[thin,<-]  (0.5,0.5) -- (1,1);
				\draw[thin,->] (0,2)  -- (0.5,1.5);
		\draw[thin]  (0.5,1.5) -- (1,1);
				\node at (-0.5,1.9) {$\chi_\beta,p$};
	\node at (-0.5,0.1) {$\bar{\chi}^\alpha,p'$};
				\draw[dashed] (1,1)  -- (2,1);
					\node at (2,0.7) {$\rho_{\mu\nu}$};
\end{tikzpicture}
\quad \quad
\begin{array}{c}
 - \frac{f_V}{2\Lambda^2} \bigg{(}i(p + p')_\nu\gamma^\mu -i(p + p')_\mu\gamma^\nu \bigg{)}
({\alpha_ d}-{\alpha_ u})
\\ \
\\ \
\\ \
\\ \
\end{array}
\eea

\bea
\begin{tikzpicture}
		\draw[thin] (0,0)  -- (0.5,0.5);
		\draw[thin,<-]  (0.5,0.5) -- (1,1);
				\draw[thin,->] (0,2)  -- (0.5,1.5);
		\draw[thin]  (0.5,1.5) -- (1,1);
				\node at (-0.5,1.9) {$\chi_\beta,p$};
	\node at (-0.5,0.1) {$\bar{\chi}^\alpha,p'$};
				\draw[dashed] (1,1)  -- (2,1);
					\node at (2,0.7) {$\omega_{\mu\nu}$};
\end{tikzpicture}
\quad \quad
\begin{array}{c}
  \frac{f_V}{2\Lambda^2} \bigg{(}i(p + p')_\nu\gamma^\mu -i(p + p')_\mu\gamma^\nu \bigg{)}
({\alpha_ d}+{\alpha_ u})
\\ \
\\ \
\\ \
\\ \
\end{array}
\eea

\bea
\begin{tikzpicture}
		\draw[thin] (0,0)  -- (0.5,0.5);
		\draw[thin,<-]  (0.5,0.5) -- (1,1);
				\draw[thin,->] (0,2)  -- (0.5,1.5);
		\draw[thin]  (0.5,1.5) -- (1,1);
				\node at (-0.5,1.9) {$\chi_\beta,p$};
	\node at (-0.5,0.1) {$\bar{\chi}^\alpha,p'$};
				\draw[dashed] (1,1)  -- (2,1);
					\node at (2,0.7) {$\phi_{\mu\nu}$};
\end{tikzpicture}
\quad \quad
\begin{array}{c}
 \sqrt{2} \frac{f_V}{2\Lambda^2} \bigg{(}i(p + p')_\nu\gamma^\mu -i(p + p')_\mu\gamma^\nu \bigg{)}
{\alpha_ s}
\\ \
\\ \
\\ \
\\ \
\end{array}
\eea

\bea
\begin{tikzpicture}
		\draw[dashed] (0,0)  -- (0.5,0.5);
		\draw[dashed]  (0.5,0.5) -- (1,1);
				\draw[dashed] (0,2)  -- (0.5,1.5);
		\draw[dashed]  (0.5,1.5) -- (1,1);
			\node at (-0.5,1.9) {$\pi, p_\pi$};
	\node at (-0.5,0.1) {$\rho_{cd}, p_\rho$};
		\node at (2,0.7) {$\omega_{ab}, p_\omega$};
					\draw[dashed] (1,1)  -- (2,1);
\end{tikzpicture}
\quad \quad
\begin{array}{c}
\frac{1}{2\sqrt{F}}\bigg{(}2h_A\varepsilon^{cdbh}(p_\pi)_h(p_\omega)^a
+ 2h_A \varepsilon^{abdh}(p_\pi)_h(p_\rho)^c
+ h_O\varepsilon^{cdgb}(p_\rho)_g(p_\pi)^a+
h_O \varepsilon^{abgd}(p_\omega)_g(p_\pi)^c\bigg{)}
\\ 
\\ \
\\ \
\\ \
\end{array}
\eea

\section{Decay Spectra at Rest}
\label{decayspectrumappendix}
We begin by finding the decay spectrum at rest for the various mesons.
For two-body decays the spectra are set by kinematics. For
a decay $A\to B+C$ the energy spectrum of $B$ is
\bea
{d\Gamma\over dE_B}=\delta \left(E_B-{m_{B}^2+m_A^2-m_{C}^2\over 2m_A} \right) .
\eea
Three body decays must be analyzed in terms of Dalitz plots,
which encode the amplitudes as a function of
the kinematic variables.

\subsubsection{$K^+(p_K)\to  \bar \ell(p_1)\nu(p_2)\pi^0(p_3)$}

The decay to $K^+(p_K)\to \bar \ell(p_1)\nu(p_2)\pi^0(p_3)$ is controlled by the matrix element
${\cal M}=(p_K+p_3)^\mu \bar{\ell}\gamma_\mu(1-\gamma_5)\nu$. The pion energy spectrum
is controlled by
\bea
\Gamma=\int dm_{12}^2 dm_{23}^2 ~\sum_{spins} |{\cal M}|^2 .
\eea
We then find that
\bea
{d\Gamma\over dm_{12}^2}\propto \int  dm_{23}^2 ~|{\cal M}|^2 &\propto&
-8m_\ell^2(m_\pi^2+m_\ell^2-m_{12}^2)-8m_K^2m_\pi^2   ((m_{23}^2)_{max}-(m_{23}^2)_{min})
\nonumber\\
&\,& +4(m_K^2+2m_\ell^2+m_\pi^2-m_{12}^2)((m_{23}^4)_{max}-(m_{23}^4)_{min})
\nonumber\\
&\,& -{8\over 3}((m_{23}^6)_{max}-(m_{23}^6)_{min}) ,
\eea
where we have defined
\bea
E_2^*={m_{12}^2-m_1^2+m_2^2\over 2m_{12}}
\qquad
E_3^*={M^2-m_{12}^2-m_3^2\over 2m_{12}} ,
\nonumber\\
&\hspace{-9cm}  m_{ij}^2 = (p_i - p_j)^2  \,,
 \nonumber \\
(m_{23}^2)_{max}=(E_2^*+E_3^*)^2-
\left(\sqrt{(E_2^*)^2-m_2^2}-\sqrt{(E_3^*)^2-m_3^2} \right)^2 ,
\nonumber\\
(m_{23}^2)_{min}=(E_2^*+E_3^*)^2-
\left(\sqrt{(E_2^*)^2-m_2^2}+\sqrt{(E_3^*)^2-m_3^2} \right)^2 .
\eea

\subsubsection{$K^+(p_K)\to \pi^0(p_1)\pi^0(p_2)\pi^+(p_3)$}

The  decay $K^+(p_K) \rightarrow \pi ^{0}(p_1)\pi
^{0}(p_2)\pi ^{+}(p_3)$ can be expressed in terms of the invariant amplitude
$\mathcal{A}$
	\cite{pdg}

\bea
|\mathcal{A}|^2 &=& \frac{d\Gamma}{ds_3 ds_2}  \propto 1
+ g \frac{(s_3-s_0)}{m_+^2} + h \left(\frac{(s_3-s_0)}{m_+^2}\right)^2
+ k\left(\frac{s_2-s_1}{m_+^2}\right)^2  ,
\nonumber\\
&=& 1 + g \frac{(s_3-s_0)}{m_+^2} + h \left(\frac{(s_3-s_0)}{m_+^2}\right)^2
+ k\left(\frac{2s_2 +s_3  -3s_0  }{m_+^2}\right)^2 ,
\eea
where
\bea
s_i=(p_K-p_i)^2
\qquad
s_0={s_1+s_2+s_3\over 3}={1\over 3}(m_K^2+m_1^2+m_2^2+m_3^2) ,
\nonumber\\
g=0.626
\qquad
h=0.052
\qquad
k=0.0054 .
\eea

The pion energy distribution is then
\bea
{d\Gamma\over dE_2}=-2m_K{d\Gamma\over ds_2}
\propto \int ds_3 ~|{\cal A}|^2 .
\eea

We then find that the energy distribution of $\pi_0$ in decays of
$K^+\rightarrow \pi ^{0}\pi
^{0}\pi ^{+}$ at rest  is
(here $E_2$ is the neutral pion energy)
\bea
{d\Gamma \over dE_{2 }} &=&
\mathcal{N}_{K^+ \to \pi^+ \pi^0 \pi^0}\sigma(s_{2}) \left( \alpha_{0} + +\alpha s_{2}
+ {\alpha_{1} \over s_{2} } + {\alpha_{2} \over s_{2}^{2}}
+ {\alpha_{3} \over s_{3}^{3}} \right) ,
\eea
where
\bea
\sigma(s_{2}) = \frac{\sqrt{\left[-2 m_+^2 \left(m_{\pi }^2+s_2\right)
+\left(m_{\pi }^2-s_2\right){}^2+m_+^4\right] \left[-2 m_{\pi }^2
 \left(m_{K_+}^2+s_2\right)+\left(m_{K_+}^2-s_2\right){}^2+m_{\pi }^4\right]}}
{2 m_+^2} ,
\eea
and
\bea
\begin{cases}
&\alpha_{0} = -g
- \frac{2}{3} h \left(1+ \frac{ m_{K_+}^2+ 2m_{\pi }^2}{ m_+^2}\right) ,\\
&\alpha =\frac{2}{3m_{+}^{2}} h ,\\
&\alpha_{1} = 2 m_+^2+ \frac{g}{3} \left(m_{K_+}^2+m_+^2+2 m_{\pi }^2\right)
+\frac{h}{9} \left(\frac{20m_{\pi}^{4} + 2m_{K_{+}}^{4} -4 m_{\pi }^2 m_{K_+}^2}
{ m_+^2}+16 m_{K_+}^2-4 m_{\pi }^2+2 m_+^2\right) ,\\
& \alpha_{2} = g \left(m_{\pi }^2 m_{K_+}^2-m_+^2 m_{K_+}^2-m_{\pi }^4
+m_+^2 m_{\pi }^2\right)+ \frac{2}{3} h \left(\frac{ m_{\pi }^4 m_{K_+}^2
-2m_{\pi}^{6}}{ m_+^2}+ 2 m_{\pi }^2 m_{K_+}^2- m_+^2 m_{K_+}^2- m_{K_+}^4
+m_{\pi }^4\right) ,\\
&\alpha_{3} = 2 h\frac{ \left(m_+^2-m_{\pi }^2\right){}^2 \left(m_{\pi }^2
-m_{K_+}^2\right){}^2}{3 m_+^2} .
\end{cases}
\eea
We find that the normalization constant by integrating both sides with respect
to $dE$ (picking up a factor of $-\frac{1}{2 m_{K_{+}}}$). The normalization
constant is given by,
\[\mathcal{N}_{K^+ \to \pi^+ \pi^0 \pi^0}=-0.00187217 \ (\gev)^3 . \]

\subsubsection{$K_L(p_K) \rightarrow \pi ^{+}(p_1)\pi
^{-}(p_2)\pi ^{0}(p_3), \pi ^{0}(p_1)\pi
^{0}(p_2)\pi ^{0}(p_3)$}

The  decay $K_L(p_K) \rightarrow \pi ^{+}(p_1)\pi
^{-}(p_2)\pi ^{0}(p_3)$ can be expressed in terms of the invariant amplitude ${\cal A}$
	\cite{pdg}
\bea
|{\cal A}(s_1,s_2,s_3)|^{2}=1+ g{(s_3-s_0)\over m_{\pi^+}^2}+ h\left({s_3-s_0\over m_{\pi^+}^2}\right)^2
+j{(s_2-s_1)\over m_{\pi^+}^2}
+k\left({s_2-s_1\over m_{\pi^+}^2}\right)^2
+f{(s_2-s_1)\over m_{\pi^+}^2}{(s_3-s_0)\over m_{\pi^+}^2}+....
\eea
Here
\bea
s_i=(p_K-p_i)^2
\qquad
s_0={s_1+s_2+s_3\over 3}={1\over 3}(m_K^2+m_1^2+m_2^2+m_3^2) ,
\nonumber\\
g=0.678
\qquad
h=0.076
\qquad
j=0.001
\qquad
f=0.0045
\qquad
k=0.0099 .
\eea
The pion energy distribution is then (here $E_3$ is the neutral pion energy)
\bea
{d\Gamma\over dE_3}
\propto \int ds_2 ~|{\cal A}|^2
\eea
We then find that the energy distribution of
 $\pi_0$ in decays of  $K_L \rightarrow \pi ^{+}\pi
^{-}\pi ^{0}$ at rest  is proportional to
\bea
{d\Gamma\over dE_3} &=&
\left[1+ g{(s_3-s_0)\over m_\pi^2}+ h({(s_3-s_0)\over m_\pi^2})^2
+(j+f{(s_3-s_0)\over m_\pi^2}){(-3s_0-s_3)\over m_\pi^2}
+k({(-3s_0-s_3)\over m_\pi^2})^2 \right]\sigma
\nonumber\\
&\,& +\left(j+f{(s_3-s_0)\over m_\pi^2}
+2k{(-3s_0-s_3)\over m_\pi^2} \right)\sigma (3s_{0}-s)
\nonumber\\
&\,& +{4\over 3}k{1\over m_\pi^4}\sigma \left({3\over 4}(3s_{0}-s)^2+{1\over 4}\sigma^2 \right) ,
\eea
where
\bea
\sigma  =\sqrt{1-\frac{4M_{\pi }^{2}}{s}}
\left( s-\left( M_{K
}+M_{\pi }\right) ^{2}\right)^{1/2} \left( s-\left( M_{K }-M_{\pi }\right)
^{2}\right)^{1/2} .
\eea

For the $K_L \to 3\pi^0$ decay, we have~\cite{pdg}
\bea
|{\cal A}|^{2} = 1+h\frac{(s_3-s_0)^2}{m_+^4} ,
\eea
where $h=0.59$. The energy distribution may then be found similarly to the calculation for
$K_L \to \pi^+\pi^-\pi^0$ above.

\subsubsection{$\omega \rightarrow \pi^0(p_1)\pi^+(p_2)\pi^-(p_3)$}
$\omega$ decays to three pions are described by a similar Dalitz plot,
which was taken from \cite{Adlarson:2016wkw}.

\section{Boosting the Decay Spectrum}
\label{boostdecayspectrumappendix}
We now describe the general procedure to obtain the boosted spectrum from the
decay spectrum at rest.

We consider
a particle of mass $m$ which
decays into a number of daughter particles, and we assume that the kinematic distribution
of the decay is known in the rest frame of the particle.  Our goal is to determine the
kinematic distribution
in the lab frame, where the parent particle is moving.
We take the parent particle to be traveling  along the $z$-axis, with an energy $E_m$,
corresponding to a Lorentz factor $\gamma=E_m/m$.
We assume that there is no correlation between the direction of the
daughter particle's momentum and the direction of the parent particle's boost.

In the CM frame, the  four-momentum of one of the daughter
particles is $(E', p'\sin\theta', 0, p'\cos\theta')$.
We are given ${dP / dE'}$  in the CM frame; i.e. the probability of obtaining in
the CM frame
a given value of the daughter particle's energy.
In the lab frame,  the four-momentum is $(E, p\sin\theta, 0, p\cos\theta)$.
We are  looking for ${dP / dE}$.

For the daughter particle in the lab frame, we have
\bea
E=\gamma(E'+p' \beta \cos\theta) .
\eea
For any given $E$, this equation has a solution for $\cos \theta$ if $E'$ lies in the range
\bea
\gamma(E-p\beta) \leq E'  \leq \gamma(E+p\beta) .
\eea

The kinematic distribution of the daughter particle in the laboratory frame is then
\bea
\frac{dP(E)}{dE} &=& \frac{1}{2} \int dE'~d\cos \theta \frac{dP(E')}{dE'} \delta
\left(E- \gamma(E' + \beta p' \cos \theta) \right) ,
\nonumber\\
&=& \frac{1}{2} \int_{E_1}^{E_2} dE' \frac{dP(E')}{dE'} {1\over p' \beta\gamma} ,
\eea
where
$E_2=\gamma(E+p\beta)$
and $E_1=\gamma(E-p\beta)$.  This formula allows us to obtain the boosted spectrum from the
decay spectrum at rest.

If we assume that the parent particle itself has a kinematic distribution in the laboratory
frame given by $dN_m / dE_m$, we then find
\bea
\frac{dP(E)}{dE} &=&\frac{1}{2} \int dE_m \frac{dN_m}{dE_m}
\int_{E_1 (E_m)}^{E_2 (E_m)} dE' \frac{dP(E')}{dE'} {1\over p' \beta \gamma} .
\eea
Moreover, if the daughter particle itself decays isotropically to some
tertiary product, one can determine kinematic distribution of this
tertiary product by simply repeating the above process, treating the daughter
particle now as the parent to the tertiary particle.

We can apply this formalism to the case of the $\pi^0$, whose dominant decay is to two photons.
In the rest frame, the photons have back-to-back momenta and
the distribution is $dP /dE'=2\delta(E'-{m_\pi\over 2})$, where
the factor of two accounts for the two photons.
We then find
\bea
\frac{dP}{dE} &=&    \int_{E \gamma (1-\beta)}^{E \gamma (1+\beta)} dE'
\delta \left(E'-\frac{m_\pi}{2} \right){1\over E' \beta\gamma}
\nonumber\\
&=& {2\over \sqrt{E_\pi^2 -m_\pi^2} } \times \left[ \theta\left(E- \frac{m_\pi}{2}
\sqrt{\frac{1-\beta}{1+\beta}} \right)
\theta\left(\frac{m_\pi}{2} \sqrt{\frac{1+\beta}{1-\beta}} -E \right)\right]
\eea
This reproduces the usual box distribution.

If the $\pi^0$ injection spectrum is given by $dN_\pi / dE_\pi$, then we
may express the photon spectrum as~\cite{Boddy:2016hbp}
\bea
  \frac{dN_\gamma}{dE_\gamma} &=&  \int_{\frac{m_\pi}{2} (\frac{2E_\gamma}
	{m_\pi}+\frac{m_\pi}{2E_\gamma})}^\infty dE_\pi
    \left[ \frac{dN_\pi}{dE_\pi} \, \frac{2}{\sqrt{E_\pi^2 - m_\pi^2}}
    \right]~
  \label{eq:GenSpectrum}
\eea
This implies that the photon spectrum is log-symmetric about $m_\pi /2$ with a global maximum at
that point.  Moreover, the spectrum decreases monotonically as the energy either increases or
decreases away from $m_\pi /2$.  We see these features in Figure~\ref{Fig:KLKS_Spectrum}.

The last thing which is needed is $dN_\pi / dE_\pi$.  This can be determined from the
procedure described above,
treating the $\pi^0$ as the daughter produced by the decay of $K_L$, $K_S$, $K^\pm$,
$\rho^\pm$ and $\omega$.  But there
is one subtlety to note.  This approach is strictly valid only if there is no
correlation between the pion boost and the
boost of the parent.  This is necessarily true if the parent is spin-0, but
need not be true if the parent is a vector meson.
But we will assume that this effect is negligible, and ignore it henceforth.

\end{document}